\documentclass[10pt,pra,letterpaper,typeset,twocolumn,superscriptaddress,showpacs,floatfix]{revtex4-2}
\usepackage{color}
\usepackage{graphicx}
\usepackage{amsmath}
\usepackage{amssymb,amsthm}
\usepackage{hyperref}
\usepackage{mathtools}

\usepackage{caption,subcaption}
\graphicspath{{pict/}{}}

\usepackage[applemac]{inputenc}

\usepackage{pgf,tikz}
\usepackage{mathrsfs}
\usepackage{bm}
\usepackage{array}
\usepackage{blkarray}

\usepackage{algorithm} 
\usepackage{algpseudocode} 
\usepackage{float}

\usepackage{hyperref}
\hypersetup{colorlinks=true,linkcolor=blue,citecolor=blue,urlcolor=blue}

\begin{document}
	
\title{Mean-squared-error-based adaptive estimation of pure quantum states and unitary transformations}
	
\author{A. Rojas} 
\email[corresponding author:]{alejarojas@udec.cl}
\affiliation{Instituto Milenio de Investigaci\'on en \'Optica y Departamento de F\'{\i}sica, Universidad Concepci\'on, casilla 160-C, Concepci\'on, Chile}

\author{L. Pereira}
\affiliation{Instituto de F\'{\i}sica Fundamental IFF-CSIC, Calle Serrano 113b, Madrid 28006, Spain.\\}

\author{S. Niklitschek}
\affiliation{Facultad de Ciencias F\'isicas y Matem\'aticas, Departamento de Estad\'istica, Universidad de Concepci\'on, casilla 160-C, Concepci\'on, Chile}

\author{A. Delgado}
\affiliation{Instituto Milenio de Investigaci\'on en \'Optica y Departamento de F\'{\i}sica, Universidad Concepci\'on, casilla 160-C, Concepci\'on, Chile}
	
\date{\today}
	
\begin{abstract}
In this article we propose a method to estimate with high accuracy pure quantum states of a single qudit. Our method is based on the minimization of the squared error between the complex probability amplitudes of the unknown state and its estimate. We show by means of numerical experiments that the estimation accuracy of the present method, which is given by the expectation of the squared error on the sample space of estimates, is state independent. Furthermore, the estimation accuracy delivered by our method is close to twice the Gill-Massar lower bound, which represents the best achievable accuracy, for all inspected dimensions. The minimization problem is solved via the concatenation of  the Complex simultaneous perturbation approximation, an iterative stochastic optimization method that works within the field of the complex numbers, and Maximum likelihood estimation, a well-known statistical inference method. This can be carried out with the help of a multi-arm interferometric array. In the case of a single qubit, a Mach-Zehnder interferometer suffices. We also show that our estimation procedure can be easily extended to estimate unknown unitary transformations acting on a single qudit. Thereby, the estimation of unitary transformations achieves a higher accuracy than that achieved by processes based on tomographic methods for mixed states.
\end{abstract}	

\pacs{Valid PACS appear here}

\maketitle

\section{Introduction}
\label{Introduction}


Recently, the estimation of unknown quantum states has been studied from the point of view of the estimation accuracy achievable by means of an ensemble of $N_T$ identically prepared copies of the state to be estimated \cite{MAHLER,STRAUPE,GUO,ADAPTIVE,STRUCHALIN,5B,CI5BB}. The ultime mixed-state estimation accuracy is given by the Gill-Massar lower bound \cite{GILL-MASSAR}, which establishes the highest possible accuracy achievable by means of separable measurements on the members of the ensemble. For instance, the mean value $\bar I$ of the Uhlmann-Josza infidelity $I=(Tr\sqrt{\sqrt{\rho}\tilde\rho\sqrt{\rho}})^2$ \cite{Josza,Uhlmann} on the set of estimates $\tilde\rho$ of the unknown state $\rho$ can be employed as a metric for the estimation accuracy. In this case the Gill-Massar lower bound becomes $\bar I\ge I_{GM}^{(mixed)}=(d^2-1)(d+1)/4N_T$ \cite{Gill-Massar}. 

Adaptive two-stage standard quantum tomography saturates $I_{GM}^{(mixed)}$ for a single qubit \cite{GUO}. In a first stage, standard quantum tomography \cite{SQT1,SQT2} is carried out on an ensemble of size $N_0$ and a first estimate $\tilde\rho_0$ is inferred. The eigenbasis of $\tilde\rho_0$ is used to construct three new mutually unbiased bases, which are then employed to perform a second stage of standard quantum tomography on an ensemble of size $N_T-N_0$. Thereafter, the acquired data is post-processed via maximum likelihood estimation \cite{MLE1,MLE2} to obtain a final estimate $\tilde\rho$. This procedure clearly requires the capacity to adapt the measurement bases and doubles the total number of measurement outcomes. Unfortunately, adaptive two-stage standard quantum tomography departs from the Gill-Massar lower bound in the case of a single qudit with $d>2$ \cite{ADAPTIVE}.

In the particular case of pure states it has been shown that a much better accuracy can be obtained. In this case, the Gill-Massar lower bound for the infidelity is $I_{GM}^{(pure)}=(d-1)/N_T$. The 5-bases based quantum tomographic method \cite{5B} produces an infidelity that lays in between $I_{GM}^{(mixed)}$ and $I_{GM}^{(pure)}$ \cite{CI5BB}. This method employs an adaptive scheme where measurements on the canonical base are employed to define four new measurement bases. The five bases determine univocally any pure state of a single qudit and allow to certify the purity assumption. 

An estimation accuracy closer to $I_{GM}^{(pure)}$ can be achieved by means of formulating the problem of quantum state determination as an optimization problem \cite{FERRIE} and solving it by means of a combination of stochastic optimization on the field of the complex numbers and maximum likelihood estimation (MLE) \cite{UTRERAS,ZAMBRANO}. In this approach the infidelity is considered a real function of complex arguments where the unknown state plays the role of a set of fixed and unknown complex parameters. This function is optimized by means of the Complex simultaneous stochastic approximation (CSPSA) method, which allows to handle non-holomorphic functions with unknown parameters. This optimization method requires the measurement of the infidelity at each iteration. The information provided by the sequence of measurements can be employed to enhance the rate of convergence of the optimization method when processed via maximum likelihood estimation.

Here, we study the estimation of pure quantum states using the mean-squared error  (MSE) as figure of merit for the accuracy. It has been theoretically proven and experimentally demonstrated that states with a small infidelity might lead to very different physical properties \cite{Benedetti,Bina,Mandarino}. Consequently, the infidelity might turn to be inadequate to assess the estimation of high-dimensional quantum systems. Therefore, it is advisable to explore other accuracy metrics. We resort to the mean-squared error mainly because it can be inferred from experimentally acquired data, it is inexpensive to compute, it is an excellent metric in the context of optimization, and it is a desirable measure in statistics and estimation theory \cite{Casella}. We first show that the squared error (SE) between the probability amplitudes of two pure quantum states of a single qudit can be measured by means of a multi-arm interferometric array. This allows us to employ the CSPSA method to optimize SE. This iterative method and 
MLE are then combined to drive a sequence of measurements in such a way that the SE rapidly decreases at each iteration. Due to the intrinsic stochasticity of CSPSA, the estimation procedure generates, for a fixed unknown state, a sample of estimates. Via Monte Carlo numerical experiments we show that the accuracy of the estimation procedure, that is, the mean of SE on the sample of estimates (or MSE), is nearly independent of the state to be estimated. Moreover, mean and median of SE agree on the sample of estimates, which indicates a symmetric distribution of estimates without outliers. Numerical simulations show that after a few iterations the estimation of unknown states enters into an asymptotic regime that follows very closely twice the Gill-Massar bound for the MSE. 

We also apply our previous result to the estimation of unknown unitary transformations. This is an important application since the successful realization of quantum information protocols and quantum devices requires the use of efficient characterization tools. Among these, the most widely employed is Quantum process tomography (QPT). This is based on a selection of probe states that undergo the process to be estimated, which is followed by the estimation of the states generated by the process. QPT has been applied to multi-qubit processors \cite{OBrien}, quantum communications channels \cite{Wang}, coherent transport in biological mechanisms \cite{Yuen-Zhou}, ion traps \cite{Riebe}, nuclear magnetic resonance \cite{Childs}, superconducting circuits \cite{Bialczak}, nitrogen-vacancy color centers \cite{Howard}, and few-photon linear-optical systems \cite{Nambu,Martini,Altepeter}. 

A quantum process is described as a completely-positive, trace preserving (CPTP) map,  which requires $d^4-d^2$ real numbers to be completely characterized. If we know, however, that the process is unitary, then the number of independent parameters can be further reduced. For instance, $d^2+d$ measurement outcomes are necessary to distinguish among unitary transformations \cite{Baldwin}. Unitarity can be certified, for example, through randomized benchmarking. 

We can try to estimate a unitary transformation $U$ with the help of the Uhlmann-Josza fidelity by applying $U$ to the elements of an orthonormal base an estimating the generated states. This procedure, however, does not allow obtaining a set of $d$ phases of the unitary transformation because the Uhlmann-Josza fidelity is insensitive to global phases. Thereby, it is necessary to increase the number of states onto which the unitary transformation acts in order to obtain the missing complex phases. Thus, the use of the Uhlmann-Josza fidelity for estimating unitary transformations becomes equivalent to the problem of estimating $d$ pure states and $d$ unknown phases. Alternatively, it is possible to define an infidelity-guided figure of merit to compare two unitary transformations $U_1$ and $U_2$. This is given by $I(U_1,U_2)=1-|Tr(U_1U_2^\dagger)|^2/d^2$ \cite{Acin1}. This can be shown to be equal to $I(U_1,U_2)=1-|\langle\psi|(U^\dagger_1\otimes{\cal I})(U_2\otimes{\cal I})|\psi\rangle|^2$, where $|\psi\rangle$ is the two-qudit maximally entangled state $\sum_k(1/\sqrt{d})|k\rangle\otimes|k\rangle$ \cite{Acin2}. Therefore, the measurement of $I(U_1, U_2) $ requires the capability of preparing maximally entangled two-qudit states and to project onto two-qudit states, which separates the estimation of states from that of unitary transformations. Instead, we apply our results on the estimation of pure states to the estimation of unitary transformations via the optimization of SE. This allows us to handle the estimation of states and unitary transformations within the same theoretical framework, avoid the use of maximally entangled two-qudit states, avoid increasing the number of measurements, and improve the estimation accuracy given a fixed number of particles interacting with the unknown unitary process. This is important, for instance, for measuring biological samples \cite{Taylor} and materials \cite{Wolfgramm} in scenarios where the number of samples (photons) must be low to avoid sample damage. Our method estimates the columns of an unknown unitary transformation separately, which after post-processing leads to an estimation accuracy for the unitary transformations close to $2d$ times the Gill-Massar bound for the MSE of a single unknown pure state. Let us note that our estimation method for states and unitary transformations, unlike the recent proposals \cite{Zhou}, measures all photons in the ensamble size independently, that is, no entanglement is used between photons. Also, most methods employed to estimate unitary transformations employ tomographic methods that are designed to estimate mixed states, which has an estimation accuracy limited by the Gill-Massar lower bound for mixed states. Thus, our estimation method for pure states leads to a clear improvement on the estimation accuracy of unitary transformations.

\section{Method}
\label{Method}
	
The optimization of a real-valued function $f({\bm x})$ of a vector ${\bm x}$ of $d$ real variables can be implemented by means of the gradient descent method. This is based on the iterative rule
\begin{equation}
\bm x_{k+1}={\bm x}_k-a_k{\bm g}(x_k),
\label{GRADIENTMETHOD}
\end{equation}
where the function ${\bm g}_k$ is the gradient $\nabla f$ and $a_k$ is a real gain coefficient. We are interested in the optimization of a real-valued function $f({\bm z},{\bm z}^*)$ with ${\bm z}$ a vector of $n$ complex variables, where the function $f({\bm z},{\bm z}^*)$ also depends on a set of unknown fix parameters. In order to optimize $f({\bm z},{\bm z}^*)$ we might attempt to employ the iterative rule
\begin{equation}
\bm z_{k+1}={\bm z}_k-a_k{\bm g}(\bm z_k).
\label{CSPSA}
\end{equation}
However, a few changes are needed. Real-valued functions of complex variables violate the Cauchy-Riemann conditions, and consequently, they lack standard complex derivative. This can be solved with the notion of Wirtinger derivatives, which are defined by \cite{WIRTINGER}
\begin{equation}
\partial_{z_i}=\frac{1}{2}(\partial_{x_i}-i\partial_{y_i})~{\rm and}~\partial_{z_i^*}=\frac{1}{2}(\partial_{x_i}+i\partial_{y_i}),
\end{equation}
where $x_i$ and $y_i$ are the real and imaginary parts of $z_i$, respectively. These derivatives exist even if the function $f$ violates the Cauchy-Riemann conditions. Minima and maxima of a real-valued function of complex variables are completely characterized by the conditions $\partial_{z_i^*}f=0~\forall~i=1,\dots,d$ or, equivalently, $\partial_{z_i}f=0~\forall~i=1,\dots,d$ \cite{BRANDWOOD,NEHARY,REMMERT}. Thereby, we can make the identification ${\bm g}=\partial_{{\bm z}^*}f$ with $\partial_{{\bm z}^*}=(\partial_{z_1^*},\dots, \partial_{z_d^*})$. Unfortunately, the function we seek to optimize depends on a set of unknown fix parameters. Thus, even when we might be able to obtain an analytical expression for the complex-valued gradient, we cannot evaluate it. In this scenario the components of the complex-valued gradient are approximated as
\begin{equation}
{g}_{k,i}=\frac{f({\bm z}_{k+},{\bm z}_{k+}^*)+\epsilon_{k,+}-f({\bm z}_{k-},{\bm z}_{k-}^*)-\epsilon_{k,-}}{2c_k{\Delta}_{k,i}^*}.
\label{ESTIMATIONCOMPLEXGRADIENT}
\end{equation}
This expression can be evaluated as long as we have access to the values of the function $f$ at the points ${\bm z}_{k\pm}={\bm z}_k\pm c_k{\bm\Delta}_k$, where $c_k$ is a real gain coefficient and the complex components ${\Delta}_{k,i}$ of vector ${\bm \Delta}_k$ provide a direction for the approximation of the complex-valued gradient at each iteration. The terms $\epsilon_{k,\pm}$ entering in Eq.\thinspace(\ref{ESTIMATIONCOMPLEXGRADIENT}) represent noise affecting the evaluations, or measurements, of the function $f$. 

Equations (\ref{CSPSA}) and (\ref{ESTIMATIONCOMPLEXGRADIENT}) are the core of the CSPSA method \cite{UTRERAS}. This allows us to optimize real-valued functions of complex variables when the complex-valued gradient cannot be evaluated. It is possible to solve this optimization problem by resorting to a real parametrization of the complex variables. However, it has been shown that CSPSA, which works on the field of the complex numbers, provides a higher rate of convergence in the estimation of pure states in comparison to similar methods working on the field of the real numbers. It has been proven that, under suitable conditions, the sequence ${\bm z}_k$ of complex vectors generated by CSPSA converges in mean to the minimizer of the function $f$. Furthermore, the approximation $\bm g$ of the complex-valued gradient is asymptotically unbiased.

The gain coefficients $a_k$ and $c_k$ entering in Eqs.\thinspace (\ref{CSPSA}) and (\ref{ESTIMATIONCOMPLEXGRADIENT}) and are given by
\begin{equation}
a_k=\frac{a}{(k+1+A)^s},~~c_k=\frac{b}{(k+1)^r}.
\end{equation}
These control the rate of convergence of CSPSA. The values of $a,A,s, b$ and $r$ are tuned to increase the rate of convergence for each function $f$. The complex coefficients $\Delta_{k,i}$  are independently and identically generated as elements in the set $\{\pm1,\pm i\}$ with uniform probability. This choice leads to a boost in the rate of convergence of the algorithm. These particular choices of gains and vector $\Delta_k$ obey the convergence conditions of CSPSA.

Let us note that the CSPSA method requires twice the evaluation of the function $f$ at each iteration. These values are employed to produce a new estimate and are thereafter discarded. This procedure thus generate a large amount of information about the function $f$ which can be also employed to accelerate the convergence of CSPSA \cite{ZAMBRANO}. For instance, this data can be processed via Maximum likelihood estimation (MLE). Given measurement settings $\textcolor{red}{S}={m_1,\dots, m_M}$ and corresponding data $D={d_1,\dots, d_M}$, maximum likelihood estimation seeks for a physical state $\rho$ that maximizes the likelihood functional
\begin{equation}
P(D|\rho,S)=\Pi^M_{i=1} P(d_i|\rho, m_i), 
\end{equation} 
with $P(d_i|\rho, m_i)$ the probability to obtain the measurement outcome $d_i$ given the state $\rho$ and measurement setting $m_i$. The optimization of the likelihood functional $P(D|\rho,S)$ requires an initial guess. This is chosen as the estimate provided by CSPSA at each iteration. The solution provided by MLE is then employed as the input estimate for the next iteration with CSPSA. Let us note that set $MS$ contains all measurements carried out until the $k$-th iteration. We employ the logarithmic multinomial likelihood functional, which is optimized employing the Nelder-Mead (or Simplex) algorithm.

To benchmark the estimation accuracy achieved by our method we will employ the Gill-Massar lower bound for the mean-squared error. 
This is a fundamental limit for the accuracy achieved in an estimation procedure. Let us consider a real-valued metric $f$ for the accuracy that is a function of the covariance matrix $\cal{C}$, the coefficients of which are given by
\begin{equation}
\mathcal{C}_{ij}(\theta)=\mathbb{E}[(\tilde\theta_i-\theta_i)(\tilde\theta_j-\theta_j)|\tilde\theta],
\end{equation}
where $\theta$ is an unknown vector and $\tilde\theta$ its estimate. The bound is the solution of the optimization problem
\begin{equation}
	\min f(\mathcal{C})
\end{equation}
with the constraint $\text{tr}(\mathcal{J}^{-1}\mathcal{I}^{N_T})=N_T(d-1)$, where $N_T$ is the total number of identical copies of the unknown state, $d$ is the dimension of the state, and  $\cal{J}$ and $\cal{I}$ are the quantum and classical Fisher matrices of the whole ensemble, respectively. Then components of the quantum Fisher matrix are given by
\begin{equation}
{\cal J}_{j,k}=\frac{1}{2}Tr(\rho[L_j,L_k]),
\end{equation}
where where $L_j$ is the symmetric logarithm derivative with respect to the parameters $S_k$ that define the state $\rho$. This derivative is implicitly defined by $\partial\rho/\partial S_j=\{\rho, L_j\}/2$. The components of the classical Fisher information matrix are given by
\begin{equation}
{\cal I}_{j,k}=\sum_m(1/p_m)(\partial p_m/\partial S_j)(\partial p_m/\partial S_k).
\end{equation}
The covariance and classical and quantum Fisher matrices are through the classical and quantum Cramer-Rao inequalities \cite{Paris} ${\cal C}\ge{\cal I}^{-1}\ge{\cal J}^{-1}$, which establish a lower limit for the Covariance matrix. In general, it is still unknown whether or not this inequality can be saturated. In the case of separable measurements, that is, measurements carried out on each individual copy of the ensemble, the quantum Cramer-Rao bound can not be attained and the fundamental limit is given by the Gill-Massar lower bound \cite{Gill-Massar}
 
If the function $f$ has the form $f(\cal{C})=\text{tr}(\cal{W}\cal{C})$ with $\cal{W}$ being a weight matrix, then the minimum is given by \cite{GUO}
\begin{equation}
	\text{tr}(\mathcal{WC})=\frac{\left[\text{tr}\left(\sqrt{\mathcal{J}^{-1/2}\mathcal{W}\mathcal{J}^{-1/2}}\right)\right]^2}{N_T(d-1)}.
\end{equation}
In the case of the mean-squared error we choose the weight matrix as the identity, in which case
we can identify the $MSE$ with $\text{tr}(\cal{C})$. For a pure state the matrix $\cal{J}$ is written as \cite{Li}
\begin{equation}
	\mathcal{J}=4\mathbb{I}_{2(d-1)\times 2(d-1)}.
\end{equation}
Thereby, the bound yields
\begin{equation}
	MSE\ge MSE_{GM}(d,N_T)=\frac{d-1}{N_T}.
\end{equation}

\section{Estimation of pure states via the optimization of the squared error}
\label{Estimation of pure states via the optimization of the squared error}


Our main aim is to obtain an estimate $|\tilde\psi\rangle = \sum_i \tilde{z}_i|i\rangle$ of an unknown quantum state $|\psi\rangle= \sum_i z_i|i\rangle$. In order to do this we consider the squared error
\begin{equation}
SE(\bm z,\bm{\tilde z})=\sum_{i=1}^{d}|z_i-\tilde z_i|^2,
\end{equation}
which is a function of the probability amplitudes $\tilde z_i$ of $|\tilde\psi\rangle$. The probability amplitudes $z_i$ of  $|\psi\rangle$ play the role of unknown fixed parameters. The unknown state $|\psi\rangle$ can be characterized as
\begin{equation}
\bm z={\rm Arg}\{\min_{\bm{\tilde z}} SE(\bm z, \bm{\tilde z})\},
\end{equation}
that is, SE achieves a global minimum when $\bm{\tilde z}=\bm z$. 

To solve the minimization of $SE(\bm z, \bm{\tilde z})$ we employ the CSPSA method. This requires the capability to obtain the value of $SE(\bm z, \bm{\tilde z})$ for all $\bm{\tilde z}$. In the simplest case of estimating the polarization state of a single photon, a Mach-Zehnder interferometer whose arms are supplemented with unitary transformations acting on the polarization degree of freedom allows us to infer the value of $SE(\bm z, \bm{\tilde z})$. In this setup, the initial state $|\psi\rangle_{in}$ of a single photon before entering the interferometer is given by $|\psi\rangle_{in}=|h\rangle_1$, where $|h\rangle_1$ describes a horizontally polarized single photon. After the interaction of the photon with the first beam splitter, the quantum state of the photon becomes $(|h\rangle_{a}+|h\rangle_{b})/\sqrt{2}$, which corresponds to an equally weighted coherent superposition of the two possible propagation paths $a$ and $b$ for the photon inside the interferometer. The polarization state of the photon changes conditional on the path, that is,
\begin{equation}
U_{a}|h\rangle_a=z_h|h\rangle_a+z_v|v\rangle_a=|\psi\rangle,
\end{equation}
which is the unknown state to be estimated, and
\begin{equation}
U_{b}|h\rangle_b=\tilde z_h|h\rangle_b+\tilde z_v|v\rangle_b=|\tilde\psi\rangle,
\end{equation}
which is the estimate of $|\psi\rangle$. Thereby, the state of the photon before the second beam splitter becomes
$[(z_h|h\rangle_{a}+z_v|v\rangle_{a})+(\tilde z_h|h\rangle_{b}+\tilde z_v|v\rangle_{b})]/\sqrt{2}$. After the second beam splitter, the state is given by
\begin{eqnarray}
|\psi\rangle_{out}&=&
\frac{1}{2}[(z_h+\tilde z_h)|h\rangle_1+(z_v+\tilde z_v)|v\rangle_1]
\nonumber\\
&+&\frac{1}{2}[(z_h-\tilde z_h)|h\rangle_2+(z_v-\tilde z_v)|v\rangle_2],
\end{eqnarray}
where the subindexes 1 and 2 indicates the output ports of the interferometer. The probability $P_2$ of detecting a photon at output port 2 is
\begin{equation}
P_2=\frac{1}{4}(|z_h-\tilde z_h|^2+|z_v-\tilde z_v|^2),
\label{P2d=2}
\end{equation}
which can be identified with the squared error of the complex probability amplitudes as
\begin{equation}
SE(\bm z, \bm{\tilde z})=4P_2.
\label{MSE-measured}
\end{equation}


Thus, in the setup above described the unitary transformation $U_a$ is employed to create the unknown polarization state $|\psi\rangle$ defined by the pair of complex probability amplitudes $(z_h, z_v)$. The unitary transformation $U_b$ is employed to generate an estimate $|\tilde\psi\rangle$, which is defined by the pair of complex probability amplitudes $(\tilde z_h, \tilde z_v)$. Equation\thinspace(\ref{MSE-measured}) indicates that the transformation $U_b$ has to be changed in such a way that no photon is detected at output port 2, in which case $|\tilde\psi\rangle=|\psi\rangle$. We employ CSPSA and MLE to drive the sequence of choices of $U_b$ toward the unknown state. Let us note that if the output ports of the interferometer are supplemented with polarizing beam splitters and single-photon detectors it is possible to measure independently the four combinations of coefficients $|z_h-\tilde z_h|^2$, $|z_v-\tilde z_v|^2$, $|z_h+\tilde z_h|^2$, and $|z_v+\tilde z_v|^2$.

The case of higher dimensions can be realized by considering a spatial qudit, that is, a qudit encoded in the propagation paths of a single photon. The initial state of the qudit is given by $|k\rangle$, where the state $|k\rangle$ describes a single photon propagating along one of several distinguishable paths $k=1,\dots,d$. On path $k$ a beam splitter transforms the state $|k\rangle$ into the superposition $(1/\sqrt{2})(|k_1\rangle+|k_2\rangle)$, where the subindexes $i=1,2$ distinguish the propagations paths at the exit ports of the beam splitter. Thereafter, on paths $k_1$ and  $k_2$ the unitary transformations $U$ and $\tilde U$ are applied, respectively. These transformations create a superposition of path states, that is, $U|k_1\rangle=\sum_kz_k|k_1\rangle$ and $\tilde U|k_2\rangle=\sum_k\tilde z_k|k_2\rangle$. This leads to the state $(1/\sqrt{2})(U|k_1\rangle+\tilde U|k_2\rangle)$. Finally, paths $k_1$ and $k_2$ for each $k=1,\dots,d$ are merged together by beam splitters, which leads to the state $(1/2)[\sum_k (z_k+\tilde{z}_k)|k_{1'}\rangle+\sum_k (z_k-\tilde{z}_k)|k_{2'}\rangle)]$ with $i=1',2'$ the output ports of each beam splitter. The probability of detecting a photon on any path $k_{2'}$ is given by
\begin{equation}
P_2=\frac{1}{4}\sum_{k=1}^d |z_k-\tilde{z}_k|^2.
\label{ANYD}
\end{equation}
Thereby, we have that
\begin{equation}
SE(\bm z, \bm{\tilde z})=4P_2,
\label{MSEANYD}
\end{equation}
which generalizes Eq.\thinspace(\ref{MSE-measured}) to the case $d>2$.

We have formulated our proposal to measure the SE in terms of bulk optics based setup. However, this proposal can easily be translated to other experimental platforms, such as, for instance, integrated quantum photonics \cite{Wang} and space-division multiplexing optical fibres \cite{Xavier,Carine}. On this platforms unitary transformations can be implemented by means of sequences composed of beam splitters and controlled phase transformations \cite{Reck,Clements}.

The main steps of the MSE-based method for estimating pure quantum states are summarized as pseudocode in Algorithm \ref{Pseudocode1} above, where we have considered the proposals to estimate the SE with the help of a multi-arm interferometer. An implementation of the pseudocode in the Python programming language can be found in the GitHub repository \cite{GITHUB}.

The minimization of SE via CSPSA requires an initial guess (or estimate) of the unknown state. Since no a priori information about the unknown state is available, the initial guess is also generated according to a uniform distribution. At each iteration, CSPSA generates the $\bm\Delta$ vector the components of which are randomly chosen. Also, at each iteration CSPSA uses the value of SE on the states $\bm{z}_{k\pm}$. According to Eq.\thinspace(\ref{MSE-measured}), the value of the SE can be inferred from a probability, which requires an ensemble of $N$  independently and identically prepared copies. Thereby, the total number of copies employed after $k$ iterations of CSPSA is given by $N_T=2Nk$. Since this ensemble is finite, the value of SE will be affected by finite statistics effects. Thus, the estimation process for a fixed unknown state has three sources of randomness: the choice of the initial guess, the choice of the $\bm\Delta$ vector, and the measurement process of SE. Thereby, each time that CSPSA is employed to obtain an estimate of a fixed unknown state $\bm{z}$, a different estimate $\bm{\tilde z}$ is generated. In this scenario the estimation accuracy for a fixed state $\bm{z}$ is given by
\begin{equation}
MSE(\bm{z})=\mathbb{E}[SE(\bm{z}, \bm{\tilde z})|\bm{\tilde z}],
\end{equation}
where the expectation is calculated over the set of all possible estimates $\bm{\tilde z}$ of $\bm{z}$. The mean-squared error can also be cast as
\begin{equation}
MSE(\bm{z})=\int \tilde p(\bm{\tilde z}) SE(\bm{z}, \bm{\tilde z}) \bm{d\tilde z},
\end{equation}
where $\tilde p(\bm{\tilde z})$ probability density function of obtaining the estimate $\bm{\tilde z}$ that characterizes the estimation procedure.

\begin{algorithm}[H]
	\caption{MSE-based estimation of pure states} 
	\begin{algorithmic}[1]
		\State Consider a known pure state $|\psi\rangle$, which is prepared on the upper arm of the interferometer by means of the transformation $U$.
		\State Choose an initial guess $|\tilde\psi_0\rangle$ and define $\tilde{z}_{0,i} = \langle i|\tilde{\psi}_{0}\rangle$.
		\State Set gain coefficients $a$, $A$, $s$, $b$ and $r$.
		\For {$k=1,\ldots, k_{max} $}
		\State Set $$a_k =\frac{a}{(k+1+A)^s},\quad c_k = \frac{b}{(k +1)^r}. $$
		\State Choose $\Delta_{k,i}$ randomly in the set $\{\pm1,\pm i \}$.
		\State Calculate $|\psi_{k\pm}\rangle=\sum_i\tilde{z}_{k\pm,i}|i\rangle/|\tilde{\bm z}_{k\pm}|$, with $\tilde{\bm z}_{k\pm} = \tilde{\bm z}_k\pm c_k{\bm \Delta_k}$.
		\State Prepare the states $|\psi_{k\pm}\rangle$ on the lower arm of the interferometer by means of the transformation $U'$
		\State Estimate experimentally the square errors $SE({\bm z},\tilde{{\bm z}}_{k\pm})$ with a sample of size $N$.
		\State Estimate the gradient as $$\tilde{g}_{k,i} = \frac{SE({\bm z}, \tilde{\bm z}_{k+})-SE({\bm z}, \tilde{\bm z}_{k-})}{ 2c_k\Delta_{k,i}^* }. $$
		\State Actualize the guess $\tilde{\bm z}_{k+1}=\tilde{\bm z}_k - a_k \tilde{\bm g}_k$.
		\State Maximize the cumulative logarithmic likelihood function using $|\phi\rangle=\sum_i\tilde{z}_{k+1,i}|i\rangle/|\tilde{\bm z}_{k+1}|$ as starting point,
		$$|\tilde{\psi}_{k+1}\rangle =\arg\max_{|\phi\rangle} \log P(D_k;|\psi\rangle,S), \quad \text{s. t.}\quad \langle \phi|\phi\rangle=1,$$
		and update the estimate as $\tilde{z}_{k+1,i}=\langle i|\tilde{\psi}_{k+1}\rangle$.
		\EndFor
	\end{algorithmic} 
	\label{Pseudocode1}
\end{algorithm}

\begin{figure}[t!]
	\centering
	\includegraphics[width=0.5\textwidth,trim= 0in 0in 0in 0in]{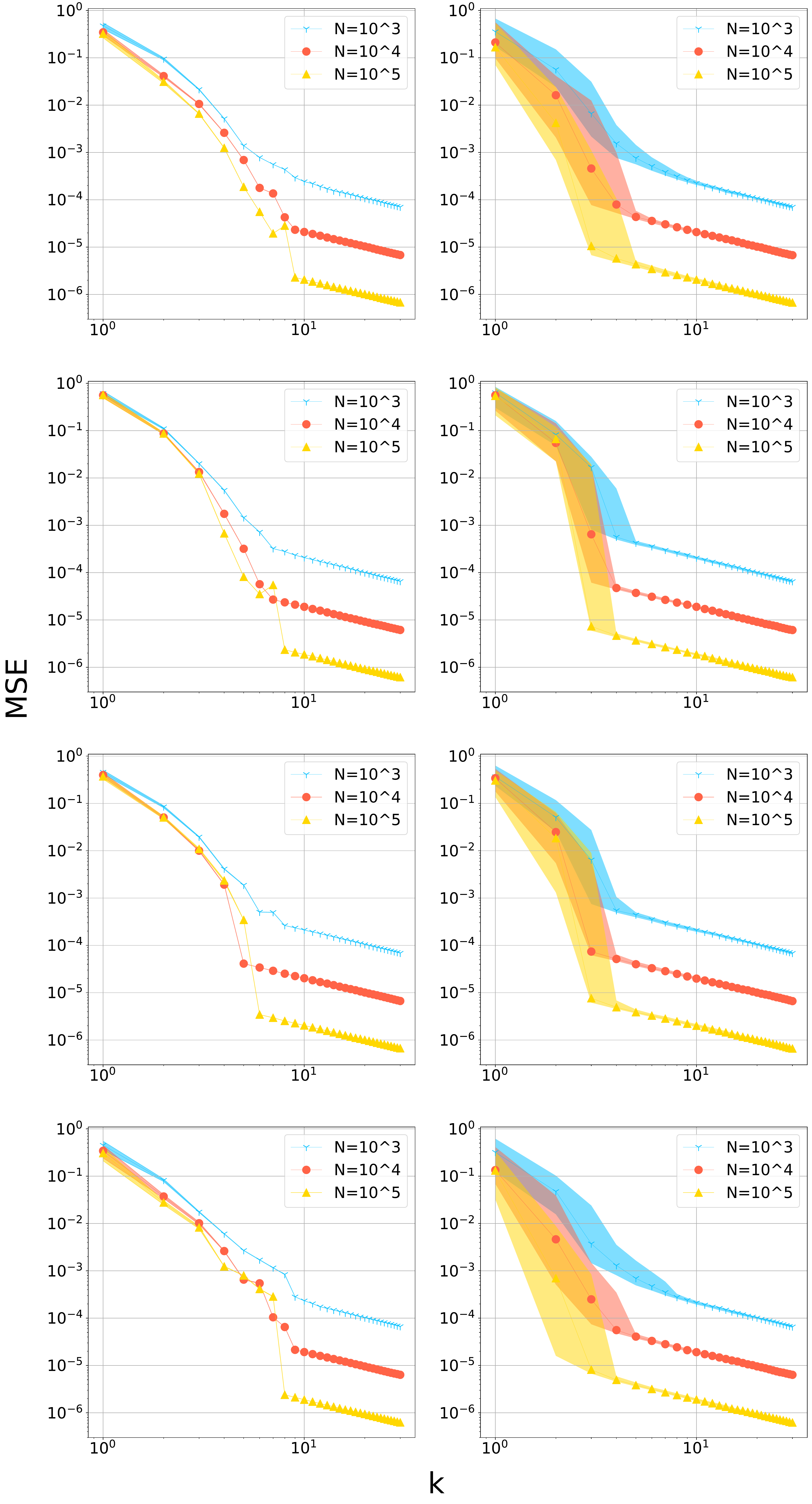}
	\caption{Left and right columns show mean and median of $SE(\bm{z}_i, \bm{\tilde z}_{i,j})$, correspondingly, with respect to $\bm{\tilde z}_{i,j}$ as a function of the number of iterations for four randomly chosen pure quantum states $\bm{z}_i$ in $d=2$ and for ensemble size $N=10^3$ (light blue down threes), $10^4$ (solid red triangles), and $10^5$ (solid yellow circles) per iteration. Shaded areas represent the corresponding interquartile range.}
	\label{Figure3}
\end{figure}

\begin{figure*}[t!]
	\centering
	\includegraphics[width=1\textwidth,trim= 0in 0in 0in 0in]{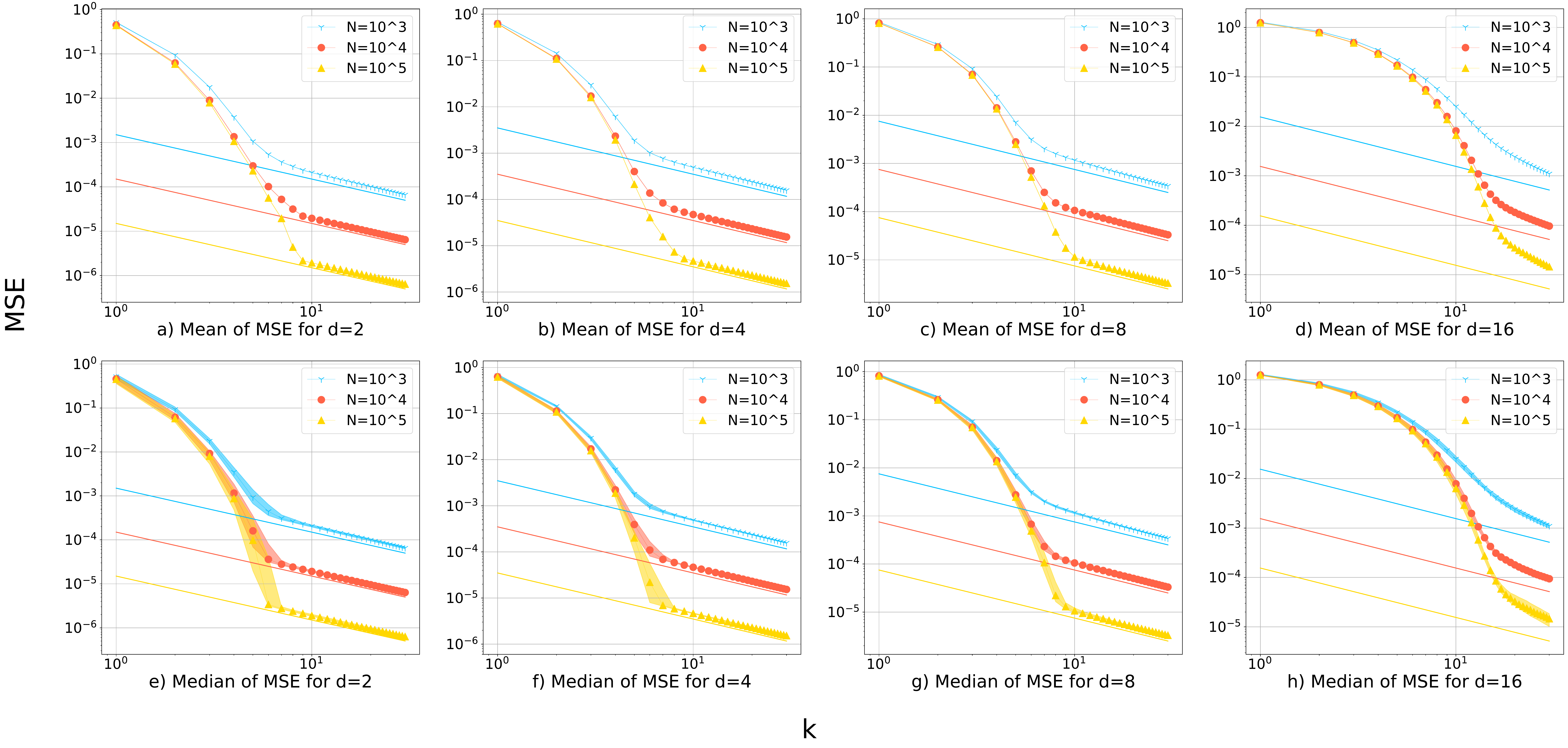}
	\caption{Mean (upper row) and median (lower row) of $MSE(\bm{z}_i)$ on $\Omega_d$ as a function of the iteration number $k$, for dimension $d$=2, 4, 8, and 16 (from left to right) and ensemble size $N=10^3$ (light blue down threes), $10^4$ (solid red triangles), and $10^5$ (solid yellow circles) per iteration. Straight lines depict the Gill-Massar lower bound $MSE_{GM}(2d,N_T)$ for the respective total ensemble size $N_T=2Nk$. Shaded areas represent interquartile range.}
	\label{Figure4}
\end{figure*}

In order to study the properties of the estimation procedure we create the set $\Omega_d=\{\bm z_i\}$ (with $i=1,\dots,m$) containing $m$ unknown pure quantum states in dimension $d$. The states in $\Omega_d$ are independently generated according to a uniform distribution. Each state in $\Omega_d$ is estimated by minimizing SE by means of CSPSA concatenated to MLE. This creates the set $\tilde\Omega_i=\{\bm{\tilde z}_{i,j}\}$ (with $j=1,\dots,n$) for each $\bm{z}_i$, which is formed by $n$ estimates $\bm{\tilde z}_{i,j}$ of $\bm{z}_i$. The estimation accuracy of $\bm{z}_i$ is given by the expectation value of $SE(\bm{z}_i, \bm{\tilde z})$ over the set of all estimates $\bm{\tilde z}$ of $\bm{z}_i$, which is approximated by the expression
\begin{equation}
MSE(\bm{z}_i)=\frac{1}{n}\sum_{j=1}^n SE(\bm{z}_i, \bm{\tilde z}_{i,j}).
\label{EstimatedMSE}
\end{equation}
The left column of Fig.\thinspace\ref{Figure3} shows $MSE(\bm{z}_i)$, calculated according to Eq.\thinspace(\ref{EstimatedMSE}), as a function of the number of iterations $k$ with ensemble size $N=10^3, 10^4$, and $10^5$, for four randomly chosen unknown states $\bm{z}_i$ in $d=2$. The four states display very similar behaviors: a rapid accuracy gain followed by an asymptotic linear regime. The latter emerges approximately after the iteration $k=6$, where $MSE(\bm{z}_i)$ reaches values close to $5\times10^{-6}$, $5\times10^{-5}$ (from top to bottom on each plot), and $5\times10^{-4}$ for $N=10^5, 10^4$, and $10^3$, respectively. The right column of Fig.\thinspace\ref{Figure3} shows the median of $SE(\bm{z}_i,\bm{\tilde z}_{i,j})$, another central tendency indicator, in the set $\tilde\Omega_i$ and the interquartile range for each one of the four states. The median also exhibits a sharp accuracy gain followed by a linear regime. However, the linear regime emerges approximately at iteration $k=3$. Before this, the median exhibits a large interquartile range, which indicates a large variation in the values of SE in the set $\tilde\Omega_i$. Once the median of SE enters into the linear regime, the interquartile range becomes extremely narrow. Thus, in the linear regime the estimation procedure leads to a very sharp distribution of values of SE. Furthermore, in each one the four inspected cases mean and median of the SE in the linear regime are almost indistinguishable. In addition, in the linear regime, the mean and median of SE are independent of the particular unknown state.

Our estimation method is based on the optimization of the SE with the help of CSPSA concatenated to MLE. Alternatively, we can reinterpret the method considering SE optimization through CSPSA to accelerate the MLE convergence rate. That is, MLE is calculated with the data obtained through SE measurements. These are chosen with the help of CSPSA in such a way that they increase the convergence of MLE towards the estimate. In this scenario, the existence of a linear regime is consistent with the efficiency and asymptotic normality of MLE estimators \cite{Lehmann}.

Figure\thinspace\ref{Figure4} displays the expectation of $MSE(\bm{z})$ over the Hilbert space of unknown states, that is,
\begin{equation}
\overline{MSE}=\mathbb{E}[MSE(\bm{z})|\bm{z}],
\end{equation}
or equivalently,
\begin{equation}
\overline{MSE}=\int p(\bm{z}) MSE(\bm{z}) \bm{dz},
\end{equation}
where $p(\bm{z})$ is the Haar-uniform probability density function for the states in the Hilbert space. This quantity is approximated as the average of $MSE(\bm{z}_i)$ over $\Omega_d$, which is given by the expression
\begin{equation}
\overline{MSE}=\frac{1}{m}\sum_{i=1}^mMSE(\bm{z}_i).
\end{equation}
Insets \ref{Figure4}(a), \ref{Figure4}(b), \ref{Figure4}(c), and \ref{Figure4}(d) show the behavior of $\overline{MSE}$ as a function of the number $k$ of iterations for dimension $d=$2, 4, 8, and 16, respectively, and for several ensemble sizes. As is apparent from these figures, the expectation of $MSE(\bm{z})$ over the Hilbert space of unknown states exhibits a rapid estimation accuracy gain followed by a linear regime. The latter arises after a number of iterations that depends on the particular dimension. In particular, the higher the dimension the more iterations are needed for the emergence of the linear regimen. The insets also depict the lower bound $MSE_{GM}(2d,N_T)$ as a function of the iteration number $k$ and the ensemble size $N$, for various dimensions. As the insets show, the estimation accuracy characteristic of our method becomes very close to $MSE_{GM}(2d,N_T)$ as $N$ increases. In fact, the accuracy of the estimates seems to be asymptotically close to $MSE_{GM}(2d,N_T)$. Insets \ref{Figure4}(e), \ref{Figure4}(f), \ref{Figure4}(g), and \ref{Figure4}(h) illustrate the median of $MSE(\bm{z_i})$ over $\Omega_d$. This exhibits a behavior similar to that of $\overline{MSE}$, but the linear regime emerges earlier. Once the optimization method enters into the linear regime, the mean and the median of $MSE(\bm{z}_i)$ reach values that cannot be distinguished. Furthermore, the interquartile range becomes extremely narrow. This indicates that in the linear regime the minimization of SE via the concatenation of CSPSA and MLE leads to an estimation procedure characterized by a state-independent MSE. Insets \ref{Figure4}(e), \ref{Figure4}(f), \ref{Figure4}(g), and \ref{Figure4}(h) also show twice the Gill-Massar lower bound for the MSE. As is apparent in all insets, in the linear regime the median of the estimation accuracy is also very close to $MSE_{GM}(2d,N_T)$.

To study the linear regime we fitted the numerical data obtained from the Monte Carlo simulations to the function $p/N_T^a$, as suggested by the observation that behavior of $\overline{MSE}$ is close to $MSE_{GM}(2d,N_T)=(2d-1)/N_T$, where $N_T=2kN$, $p=2d-1$ and $a=1$. The best fits for the values of $p$ and $a$ are shown in Table \ref{table:fit}, where two sets of $p$ and $a$ values are indicated for each dimension $d$ and ensemble size $N$. The first set of values is obtained fitting data from iteration $k=10$ until $k=45$. The second set of values is obtained fitting data from iteration $k=46$ until $k=100$. With the exception of the first dataset for $d=16$, Table \ref{table:fit} indicates that the value of $a$ are in the range $[0.99, 1.1]$ with an average value $\bar a$ to 1.01. The anomalous behavior of the first dataset for $d=16$ can be attributed to the fact that for $d=16$ more than 10 iterations are required for the onset of the linear asymptotic regime. The first dataset exhibits values of $p$ that are larger than the values of $p$ of the second dataset. In both datasets, the values of $p$ decrease with an increase in the ensemble size $N$. This indicates a decrease in the rate at which the algorithm approaches the minimizer. Finally, the values of $p$ for $N=10^4$ and $10^5$ in the second dataset are close to the value of $2d$, specially the latter. Thus, for larger values of $N$ and $k$, we can approximate $\overline{MSE}$ as
\begin{equation}
\overline{MSE}=\frac{2d+\alpha}{2kN},
\end{equation}
where $\alpha$ is a small quantity in comparison to $2d$.

\begin{table}[t]
\centering
\begin{tabular}{|l |l l |l l |l l|}
\hline\hline
& $N=10^3$& & $N=10^4$ & & $N=10^5$ &\\ [0.5ex] 
\hline
d=2 & p=6.34 &a=1.03 & p=9.97 &a=1.03 & p=6.52 &a=1.02 \\
       & p=7.41 &a=1.04 & p=4.21 &a=0.99 & p=3.57 &a=0.99 \\
d=4 & p=10.56 &a=1.00 & p=9.12 &a=0.99 & p=8.77 &a=0.99 \\
       & p=10.18 &a=1.00 & p=8.12 &a=0.99 & p=8.50 &a=0.99 \\
d=8 & p=56.69 &a=1.09 & p=25.76 &a=1.02 & p=19.66 &a=0.99 \\
       & p=21.30 &a=1.00 & p=20.16 &a=1.00 & p=17.41 &a=0.99 \\
d=16 & p=4.80 &a=2.21 & p=1.18 &a=2.09 & p=8.84 &a=2.93 \\
         & p=147.62 &a=1.10 & p=39.13 &a=0.99 & p=33.87 &a=0.99 \\[1ex]
\hline
\end{tabular}
\caption{Fit $\overline{MSE}=p/(2kN)^a$ in asymptotic regime. $MSE_{GM}(2d,2kN)$ is obtained with $p=2d-1$ and $a=1$. For $d$ and $N$ fixed, the first pair $(p,a)$ is obtained fitting iterations from $k=10$ until $k=45$. The second pair is obtained fitting iterations from $k=46$ until $k=100$.}
\label{table:fit}
\end{table}

\begin{figure}[t]
	\centering
	\includegraphics[width=0.5\textwidth,trim= 0in 0in 0in 0in]{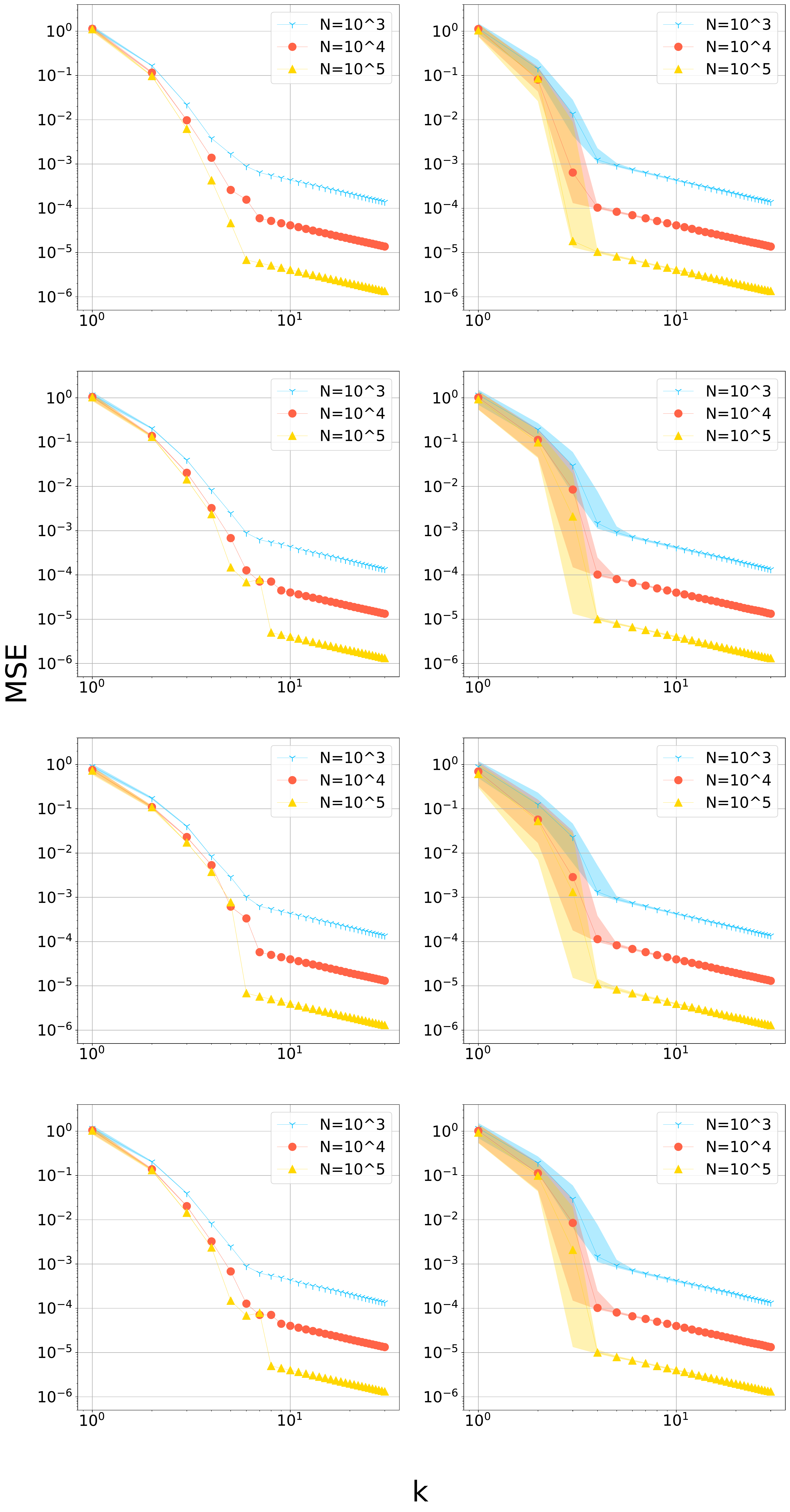}
	\caption{Right and left columns show the mean and median of $SE(\bm{\tilde z})$, respectively, as a function of the number of iterations for four randomly chosen unitary transformations in $d=2$ with ensemble size $N=10^3$ (light blue down threes), $10^4$ (solid red triangles), and $10^5$ (solid yellow circles) per iteration. Shaded areas represent the corresponding interquartile range.}
\label{Figure5}
\end{figure}

\section{Estimation of unitary transformations via the optimization of the squared error}
\label{Estimation of unitary transformations via the optimization of the squared error}

The estimation of processes acting on quantum states is a much more demanding problem than the estimation of quantum states. For instance, the estimation of an unknown process acting onto a single qudit requires the characterization of $d^4-d^2$ real parameters \cite{Nielsen}. In the case of a unitary transformation, only $d^2$ parameters must be determined. 

In general, the estimation of a quantum process is carried out by carefully choosing a set of states, letting the process act on them, and reconstructing the output states by means of a quantum tomographic method \cite{Baldwin}. We will employ this strategy to estimate an unknown unitary transformation $U$. This is suggested by Eq.\thinspace(\ref{ANYD}), which can be cast in the form
\begin{equation}
P_2=\frac{1}{4}\sum_{j=1}^d |U_{j,k}-\tilde{U}_{j,k}|^2.
\end{equation}
Thus, the probability $P_2$ is proportional to the squared error between the k-th columns of the matrices $U$ and $\tilde U$, where $k$ is controlled by the initial path state $|k\rangle$ followed by the single photon. Clearly, we can reconstruct each column of $U$ by minimizing the squared error by CSPSA and MLE. After estimating all columns of $U$ independently, we obtain an estimate $\tilde U$ of $U$.

\begin{figure}[t]
	\centering
	\includegraphics[width=0.5\textwidth,trim= 0in 0in 0in 0in]{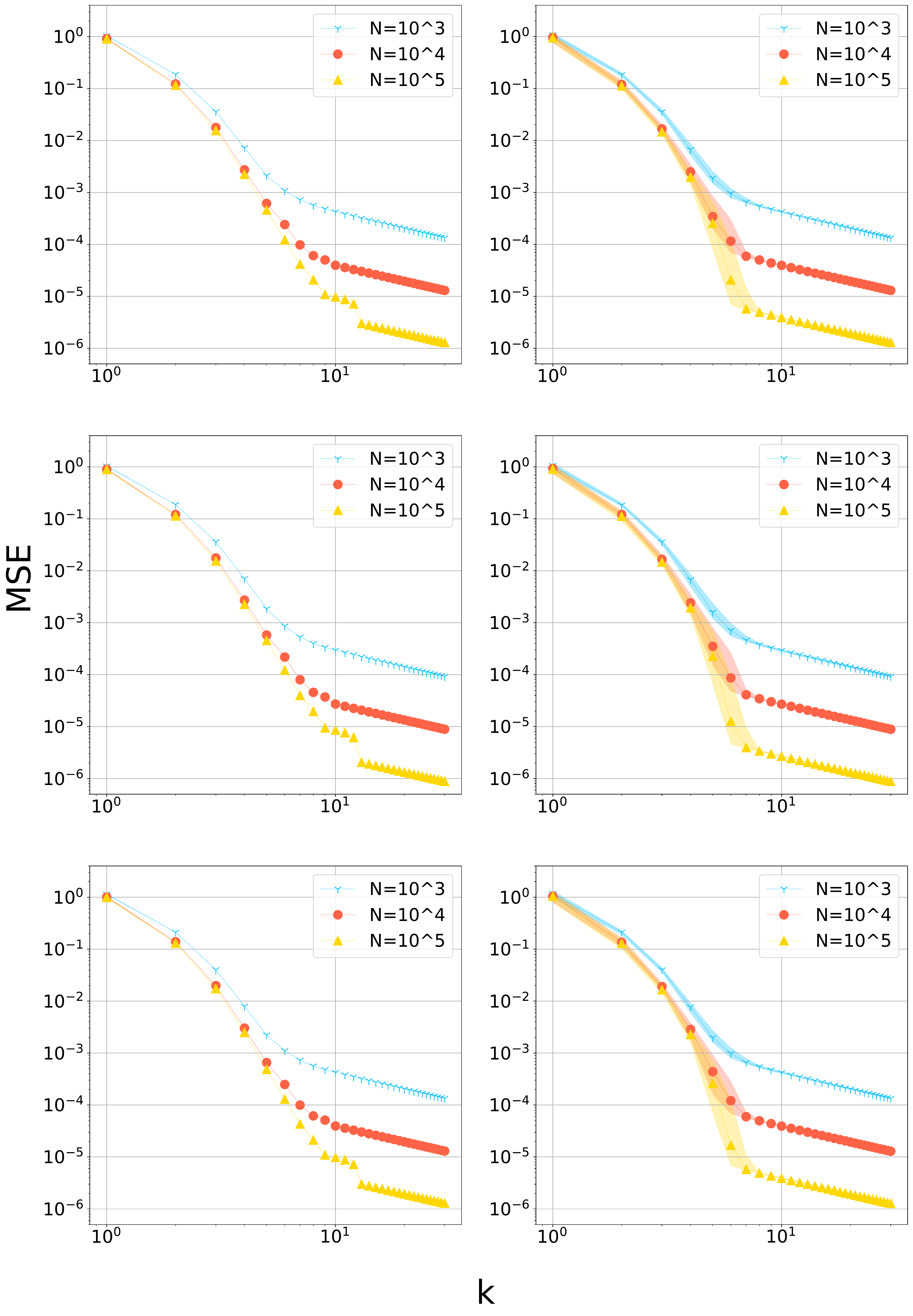}
	\caption{Mean (left column) and median (right column) of $MSE$ for randomly generated unitary transformations as a function of the iteration number $k$ for dimension $d$=2 and ensemble size $N=10^3$ (light blue down threes), $10^4$ (solid red triangles), and $10^5$ (solid yellow circles) per iteration. Shaded areas represent interquartile range. From the top row to the bottom row: estimates $\tilde U$ provided by CSPSA, estimates $\tilde U_c$ provided by CSPSA projected to the closest unitary transformation at each iteration, and estimates $\tilde U_{gs}$ provided by CSPSA post-processed with the Gramm-Schmidt orthogonalization procedure at each iteration.}
\label{Figure6}
\end{figure}

However, the present estimation method cannot guarantee that the estimate $\tilde U$ is really unitary. We consider two methods to obtain a unitary estimate. We can obtain a unitary estimate from $\tilde U$ by means of the expression \cite{Keller}
\begin{equation}
\tilde U_c=\tilde U(\tilde U\tilde U^\dagger)^{-1/2},
\end{equation}
which is the closest unitary operator to the transformation $\tilde U$. In order to quantify how close $\tilde U_c$ is from $U$ we employ the Hilbert-Schmidt distance $D(U,\tilde U_c)=Tr[(U-\tilde U_c)(U-\tilde U_c)^\dagger]$. Another method to generate a unitary estimate $\tilde U_{gs}$ consists in the application of the Gramm-Schmidt orthogonalization procedure to the columns $\tilde U_{j,k}$.

Figure\thinspace\ref{Figure5} displays the mean and median of $SE(\bm{\tilde z})$ for the estimate $\tilde U$ for four randomly chosen unitary transformations as a function of the number of iterations $k$ for different ensemble sizes and $d=2$. The overall behavior of the mean-squared error for the unitary transformations resembles very closely the case of the estimation of pure states, that is, a fast decrease of the MSE followed by a linear regime. In the case of the mean MSE the linear regime arises at a higher number of iterations when compared to the case of estimating pure states. In the case of the median, the linear regime emerges at a similar number of iterations as in the case of the median MSE for pure states. After 10 iterations, Fig.\thinspace\ref{Figure5} shows that the mean and median MSE achieve similar values, which are half order of magnitude higher than the case of estimating pure states. This entail a loss of accuracy when estimating unitary transformations with a method designed to estimate pure states. This is, however, not dramatic. After 10 iterations the MSE reaches values of the order of $0.5\times10^{-3}$, $0.5\times10^{-4}$ and $0.3\times10^{-5}$ for ensamble sizes of $10^3$, $10^4$ and $10^5$, correspondingly.

\begin{figure}[t]
	\centering
	\includegraphics[width=0.5\textwidth,trim= 0in 0in 0in 0in]{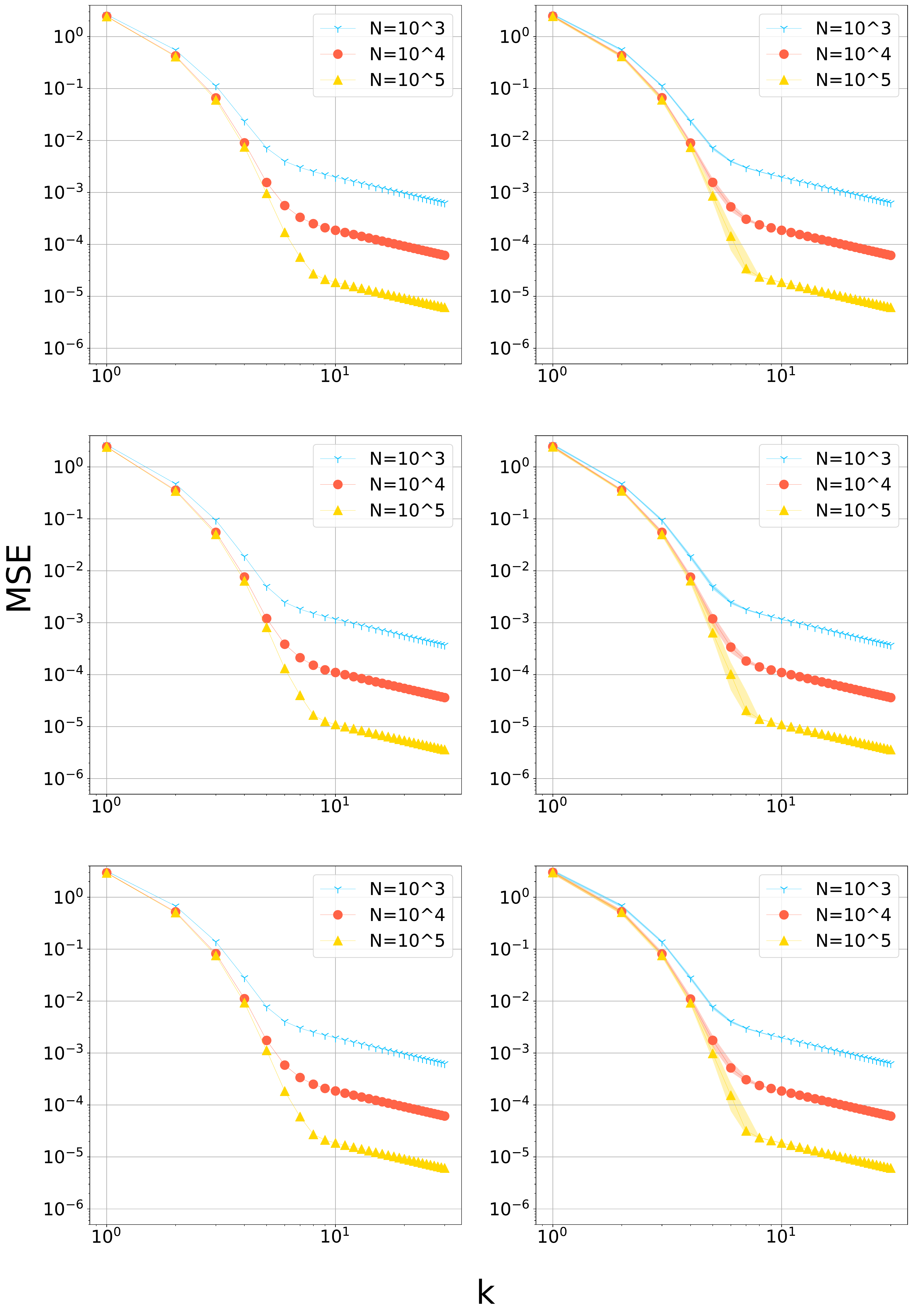}
	\caption{Mean (left column) and median (right column) of $MSE$ for randomly generated unitary transformations as a function of the iteration number $k$ for dimension $d$=4 and ensemble size $N=10^3$ (light blue down threes), $10^4$ (solid red triangles), and $10^5$ (solid yellow circles) per iteration. Shaded areas represent interquartile range. From the top row to the bottom row: estimates $\tilde U$ provided by CSPSA, estimates $\tilde U_c$ provided by CSPSA projected to the closest unitary transformation at each iteration, and estimates $\tilde U_{gs}$ provided by CSPSA post-processed with the Gramm-Schmidt orthogonalization procedure at each iteration.}
\label{Figure7}
\end{figure}

\begin{figure}[t]
	\centering
	\includegraphics[width=0.5\textwidth,trim= 0in 0in 0in 0in]{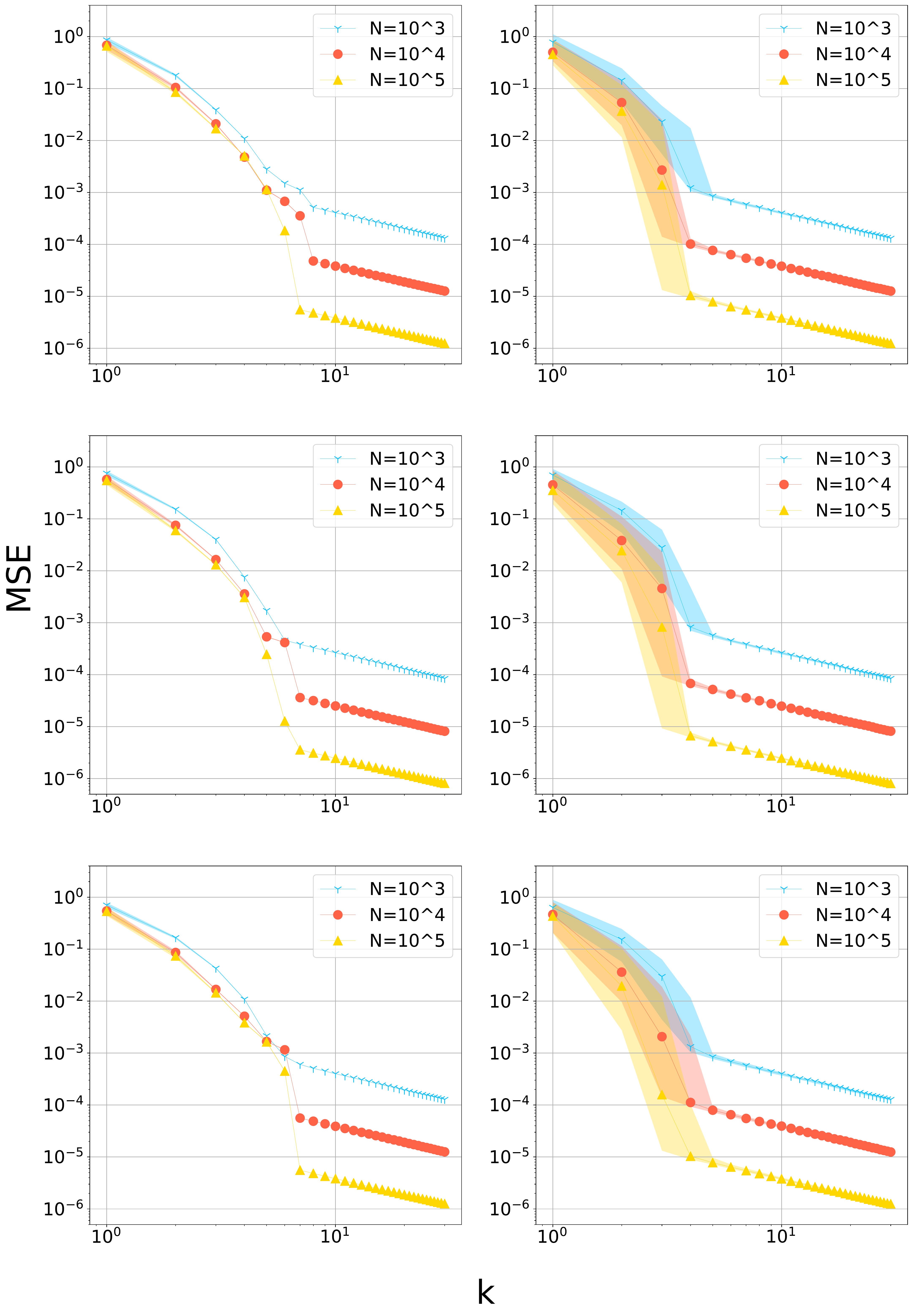}
	\caption{Mean (left column) and median (right column) of $MSE$ for randomly generated unitary transformations as a function of the iteration number $k$ for dimension $d$=2 and ensemble size $N=10^3$ (light blue down threes), $10^4$ (solid red triangles), and $10^5$ (solid yellow circles) per iteration. Shaded areas represent interquartile range. From the top row to the bottom row: estimates provided by CSPSA, estimates provided by CSPSA updated at each iteration by projection to the closest unitary transformation, and estimates provided by CSPSA updated at each iteration with the Gramm-Schmidt orthogonalization procedure.}
\label{Figure8}
\end{figure}

Figure\thinspace\ref{Figure6} shows the mean and median mean-squared error achieved in process of estimating unknown unitary transformations acting onto a 2-dimensional quantum system. The left (right) column exhibits the mean (median) achieved with the estimates $\tilde U$, $\tilde U_c$, and $\tilde U_{gs}$ from top to bottom, respectively. The typical behavior of a rapidly increasing estimation accuracy followed by a linear regimen is clearly present. This exhibits in the linear regime a mean and a median that cannot be distinguished from each other and an extremely narrow interquartile range, which indicates that after 10 iterations all unitary transformations are estimated with the same accuracy. This is almost twice the accuracy obtained in estimating a 2-dimensional pure state, as expected. The three estimates lead to very similar accuracies, but the estimate $\tilde U_c$ generates a marginally better performance. Figure\thinspace\ref{Figure7} exhibits similar results in the case $d=4$.

\begin{figure}[t]
	\centering
	\includegraphics[width=0.5\textwidth,trim= 0in 0in 0in 0in]{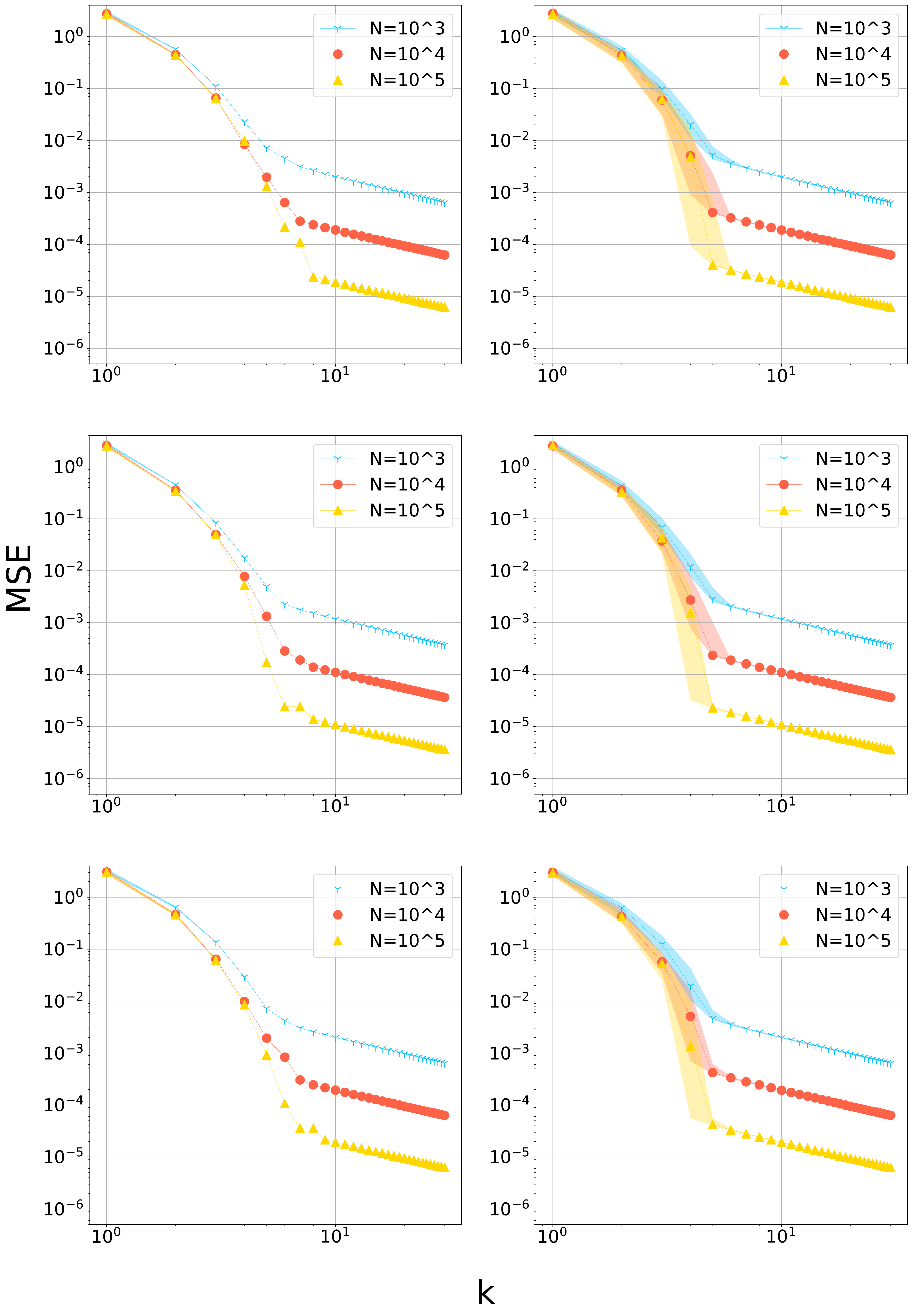}
	\caption{Mean (left column) and median (right column) of $MSE$ for randomly generated unitary transformations as a function of the iteration number $k$ for dimension $d$=4 and ensemble size $N=10^3$ (light blue down threes), $10^4$ (solid red triangles), and $10^5$ (solid yellow circles) per iteration. Shaded areas represent interquartile range. Estimates are updated after each iteration. From the top row to the bottom row: estimates provided by CSPSA, estimates provided by CSPSA updated at each iteration by projection to the closest unitary transformation, and estimates provided by CSPSA updated at each iteration with the Gramm-Schmidt orthogonalization procedure.}
\label{Figure9}
\end{figure}

In the previous simulations we have considered that after each iteration the estimate of $U$ is post-processed with the help of the Gram-Schmidt orthogonalization procedure or the projection onto the closest unitary transformation. This information has not been employed to modify the estimate along the sequence of iterations. However, we can estimate the columns of the unitary transformation, obtain a non-unitary estimate of the unitary, and generate from it a unitary estimate. Thereafter, the columns of this unitary estimate are employed as guesses for the next iteration round. Figure\thinspace\ref{Figure8} shows the effect on the estimation quality of performing such an update in the case of $d=2$. The left (right) column exhibits the mean (median) achieved with the estimate $\tilde U$ (for comparison purpose), the estimate $\tilde U_c$ updated by means of the projection onto the set of unitary transformations, and the estimate $\tilde U_{gs}$ updated by means of the Gram-Schmidt orthogonalization procedure from top to bottom, respectively. As it is apparent from this figure, the update of the estimates leads to a modest increase in estimation accuracy. However, in dimension $d=4$ there is a significative improvement in the estimation accuracy. This is illustrated in Fig.\thinspace\ref{Figure9}, where an improvement in half order of magnitude is achieved in comparison to Fig.\thinspace\ref{Figure7}, where no update of the estimates is performed. Nevertheless, the convergence rate to the linear regime is reduced by a few iterations.

In Algorithm \ref{Pseudocode2} we present a basic pseudocode for implementing the MSE-based estimation of unitary transformations. An implementation of the pseudocode in the Python programming language can be found in the GitHub repository \cite{GITHUB}. We consider the different choices for the post-processing of the estimates. According to option 1 we project the possibly non-unitary estimate onto the set of unitary transformations and evaluate the infidelity. These projections are used in option 2 to provide a better update of the estimate.

Since each of the $d$ columns of $U$ is estimated with an accuracy close to $(2d+\alpha)/N_T$, where $N_T$ is the total ensamble size used in the estimation of each column, we have that the unitary transformations are estimated with an accuracy $MSE(U)$ given approximately by
\begin{equation}
MSE(U)\approx\frac{d(2d+\alpha)}{N_T}.
\end{equation}
Thereby, the estimation accuracy of our procedure becomes
\begin{equation}
MSE(U)\approx\frac{d^2(2d+\alpha)}{N^*_T},
\end{equation}
where $N^*_T=dN_T$ is the total number of copies used in the estimation of all $d$ columns of $U$.

It is possible to follow a different approach to the estimation of a unitary transformation. The multi-arm interferometer can be feed with a maximally mixed state $I/d$. In this case the detection of photons at the output ports leads to the SE of all the elements of the unknown unitary transformation. This, however, requires the solution of the MLE problem considering all elements of a density matrix. This procedure increases the computational cost of the estimation process and entails a reduction of the estimation accuracy, which becomes more severe in higher dimensions. 

\begin{algorithm}[H]
	\caption{MSE-based estimation of unitary transformations} 
	\begin{algorithmic}[1]
		\State Consider a known unitary transformation $U$ on the upper arm of the interferometer.
		\State Choose initial estimate $\tilde U_0$ and define $\tilde{z}_{0,i}^{j} = \tilde{U}_{0,ij}$.
		\State Set gain coefficients $a$, $A$, $s$, $b$ and $r$. 
		
		\For {$k=1,\ldots, k_{max} $}
		\State Set $$a_k =\frac{a}{(k+1+A)^s},\quad c_k = \frac{b}{(k +1)^r}. $$
		\For {$j=1,\dots,d$}
		\State Choose $\Delta_{k,i}^j$ randomly in the set $\{\pm1,\pm i \}$.
		\State Calculate $|\psi_{k\pm}^j\rangle=\sum_i\tilde{z}_{k\pm,i}^j|i\rangle/|\tilde{\bm z}_{k\pm}^j|$, with $\tilde{\bm z}_{k\pm}^j = \tilde{\bm z}_k\pm c_k{\bm \Delta_k}$.
		\State Feed the interferometer in the mode $|j\rangle$.
		\State Prepare the states $|\psi_{k\pm}^j\rangle$ on the lower arm of the interferometer with the transformation $U'$.
		\State Estimate experimentally the square errors $SE({\bm z^j},\tilde{{\bm z}}^j_{k\pm})$ of the $j$-column of $U$ with a sample of size $N$.
		\State Estimate the gradient as $$\tilde{g}_{k,i}^j = \frac{SE({\bm z}^j, \tilde{\bm z}^j_{k+})-SE({\bm z}^j, \tilde{\bm z}^j_{k-})}{ 2c_k\Delta_{k,i}^* }. $$
		\State Actualize the guess $\tilde{\bm z}^j_{k+1}=\tilde{\bm z}^j_k - a_k \tilde{\bm g}^j_k$.
		\State Maximize the cumulative Likelihood function using $|\phi^j\rangle=\sum_i\tilde{z}^j_{k+1,i}|i\rangle/|\tilde{\bm z}^j_{k+1}| $ as starting point,
		$$|\tilde{\psi}_{k+1}^j\rangle =\arg\max_{|\phi\rangle} \log P(D_k^j;|U|j\rangle,S), \quad \text{s. t.}\quad \langle \phi|\phi\rangle=1,$$
		and update the estimate as $\tilde{z}_{k+1,i}^j=\langle i|\tilde{\psi}_{k+1}^j\rangle$.
		\EndFor
		\State The estimated unitary matrix is $\tilde{U}_{k+1,ij} =  \tilde{z}_{k+1,i}^j$.
		\State \textbf{Option 1}: In order to guarantee the unitarity of the estimator, we consider two postprocessing methods:
		\begin{itemize}
			\item $\tilde{U}_c$: Project $\tilde{U}$ into its closest unitary matrix.
			\item $\tilde{U}_{gs}$: Apply the Gram-Schmidt procedure to the columns of $\tilde{U}_{k+1}$.
		\end{itemize}
		\State \textbf{Option 2}: Re-update the estimates $\{\tilde{\bm z }_{k+1}^j\}$ with the postprocessed unitary matrix, $$\tilde{z}_{k+1,i}^j = (\tilde{U}_c)_{k+1,ij}\qquad\text{or}\qquad \tilde{z}_{k+1,i}^j = (\tilde{U}_{gs})_{k+1,ij}.$$ 
		\EndFor
	\end{algorithmic} 
	\label{Pseudocode2}
\end{algorithm}

\section{Conclusions}
\label{Conclusions}

We have studied the estimation of pure quantum states employing the mean-squared error as accuracy metric. We have presented two setups within reach of actual experimental techniques, one for a polarization qubit and other for a path-encoded qudit, that allow one to measure the squared error. The mean-squared error arises as a sampling of the squared error for a fixed unknown state. The estimates of an unknown state are obtained by a combination of stochastic optimization on the field of the complex numbers and maximum likelihood estimation. The estimation of an unknown state is formulated as the minimization of the infidelity with respecto to a known state. This optimization problem is solved by means of CSPSA, which iteratively drives a sequence of measurements in such a way that the infidelity approaches zero. The rate of convergence of CPSA is increased by refining the estimates with the help of maximum likelihood estimation applied onto the total sequence of measurement results.

Monte Carlo numerical experiments show that the accuracy achieved in the estimation of a fixed unknown state by the combination of CSPSA and MLE exhibits, as a function of the number of iterations, two clearly defined regimes: a fast decrease followed by a lineal behavior. The mean and median squared error exhibit very close values, which indicates the absence of outliers. The median squared error enters the lineal regime a few iterations before than the mean-squared error. In addition, numerical experiments in a wide range of dimension and ensemble size indicate that the achieved estimation accuracy is nearly state independent. Furthermore, the estimation accuracy reaches very close values to twice the Gill-Massar lower bound for the mean-squared error. 


We have also extended the estimation of pure states to the estimation of unitary transformations. In first place we hace considered the estimation by simply estimating each column of a unitary transformation, which leads to an accuracy similar to obtained in the estimation of pure states. However, in this case it is not possible to guaranty that the estimates fulfill the condition of unitarity. In order to accomplish this, the estimates are projected onto the closest unitary transformation or the estimated columns undergo the Gram-Schmidt orthogonalization procedure. In both cases there is an improvement in the estimation accuracy. A much more significative improvement is obtained when incorporating the projection onto the closest unitary transformation or the Gram-Schmidt orthogonalization procedure to the iteration process that delivers the estimates. The present method also inherits several of the characteristics exhibited in the estimation of states, that is, it is independent of the unitary to be estimated, exhibits very close values of mean and median accuracy, and a very narrow interquartile range.

%
%

\acknowledgments{This work was supported by ANID -- Millennium Science Initiative Program -- ICN17$_-$012. AD was supported by FONDECYT Grant 1180558. LP was supported by ANID-PFCHA/DOCTORADO-BECAS-CHILE/2019-72200275.}

\end{document}